\documentclass[onecolumn,11pt]{article}
\usepackage{amsmath,amssymb}
\usepackage[dvipdfmx]{graphicx}
\usepackage{setspace,atbegshi,mediabb,color}
\usepackage{cite}
\usepackage{ascmac}
\newdimen\tbaselineshift
\usepackage{framed}
\usepackage{fancybox}
\AtBeginShipoutFirst{}
\usepackage{authblk}

\textwidth=\paperwidth
\advance \textwidth by -1.5in
\textheight=\paperheight
\advance \textheight by -2.2in
\def\ket#1{|#1\rangle}

\newcommand{\rA}{{\mathrm{A}}}
\newcommand{\rB}{{\mathrm{B}}}
\newcommand{\rC}{{\mathrm{C}}}
\newcommand{\rD}{{\mathrm{D}}}

\newcommand{\tr}{{\mathrm{tr}}}

\newcommand{\rX}{{\mathrm{X}}}
\newcommand{\rY}{{\mathrm{Y}}}
\newcommand{\rZ}{{\mathrm{Z}}}

\begin{document}

\title{Scrambling of Quantum Information in Quantum Many-Body Systems}

\author[1]{Eiki Iyoda}
\author[1]{Takahiro Sagawa}
\affil[1]{{\small Department of Applied Physics, The University of Tokyo, 7-3-1 Hongo, Bunkyo-ku, Tokyo 113-8656, Japan}}

\maketitle

\begin{abstract} 
We systematically investigate scrambling (or delocalizing) processes of quantum information encoded in quantum many-body systems by using numerical exact diagonalization. As a measure of scrambling, we adopt the tripartite mutual information (TMI) that becomes negative when quantum information is delocalized. We clarify that scrambling is an independent property of integrability of Hamiltonians; TMI can be negative or positive for both integrable and non-integrable systems. This implies that scrambling is a separate concept from conventional quantum chaos characterized by non-integrability.
Furthermore, we calculate TMI in the Sachdev-Ye-Kitaev (SYK) model, a fermionic toy model of quantum gravity. We find that disorder does not make scrambling slower but makes it smoother in the SYK model, in contrast to many-body localization (MBL) in spin chains.
\end{abstract}


\textit{Introduction.}
Whether an isolated system thermalizes or not is a fundamental issue in statistical mechanics, which is related to non-integrability of Hamiltonians. In classical systems, thermalization has been discussed in terms of ergodicity of chaotic systems~\cite{Gallavotti1999}. In quantum systems, a counterpart of classical chaos is not immediately obvious, because the Schr\"odinger equation is linear. Nevertheless, it has been established that there are some indicators of chaotic behaviors in quantum systems, such as the level statistics of Hamiltonians~\cite{Jacquid1997,Santos2010,Stockmann1999} and decay of the Loschmidt echo~\cite{Gorin2006,Goussev2012}. More recently, the eigenstate-thermalization hypothesis (ETH)~\cite{Srednicki1994,Rigol2008,Biroli2010,DAlessio2016,Iyoda2016} has attracted attention as another indicator of quantum chaos in many-body systems, which states that even a single energy eigenstate is thermal. All these indicators of quantum chaos are directly related to integrability of Hamiltonians; non-integrable quantum systems exhibit chaos.
Such a chaotic behavior in isolated quantum systems is 
also a topic of active researches in real experiments with ultracold atoms~\cite{Kinoshita2006,Gring2012,Trotzky2012}, trapped ions~\cite{Clos2016}, NMR~\cite{Gorin2006}, and superconducting qubits~\cite{Neill2016}.

In order to 
investigate ``chaotic" properties of quantum many-body systems beyond the conventional concept of quantum chaos,
it is significant to focus on  dynamics of quantum information encoded in quantum many-body systems.
How does locally-encoded quantum information spread out over the entire system by unitary dynamics?
Such delocalization of quantum information is referred to as {\it scrambling}~\cite{Hayden2007,Sekino2008,Shenker2014,Maldacena2016Scramble,Hosur2016}.
Investigating scrambling is important 
not only for understanding relaxation dynamics of experimental systems at hand,
but also in terms of information paradox of black holes~\cite{Hayden2007}, where it has been argued that black holes are the fastest scramblers in the universe~\cite{Sekino2008}.
However, 
the fundamental relationship between scrambling and conventional quantum chaos has not been comprehensively understood.

Scrambling can be quantified by the tripartite mutual information (TMI)~\cite{Hosur2016,Cerf1998}, which becomes negative if quantum information is scrambled. There is also another measure of scrambling, named the out-of-time-ordered correlator (OTOC)~\cite{Kitaev2015,Hosur2016,Maldacena2016Scramble,Aleiner2016,Haehl2017,Roberts2016,Caputa2016,Kukuljan2017,Rozenbaum2017,Hashimoto2017}.
It has been argued that the decay rate of OTOC is connected to the Lyapunov exponent in the semiclassical limit~\cite{Kitaev2015}.
TMI and OTOC capture essentially the same feature of scrambling~\cite{Hosur2016},
where OTOC depends on a choice of observables but TMI does not.
In the context of the holographic theory of quantum gravity,
TMI is shown negative~\cite{Hayden2013} if the Ryu-Takayanagi formula~\cite{Ryu2006} is applied, suggesting that gravity has a scrambling property.
This is consistent with fast scrambling in the Sachdev-Ye-Kitaev (SYK) model~\cite{Sachdev1993,Kitaev2015,Sachdev2015,Michel2016,Fu2016,Maldacena2016SYK,Danshita2016,Cotler2016,Jian2017}, a toy model of a quantum black hole.
Then, a natural question raised is to what extent such a property of quantum gravity is intrinsic to gravity or can be valid for general quantum many-body systems.

In this Letter,
we perform systematic numerical calculations of real-time dynamics of TMI in quantum many-body systems under unitary dynamics, by using exact diagonalization of Hamiltonians.
We consider a small system (say, a qubit) and a quantum many-body system (say, a spin chain). 
The information of the small system is initially encoded in the many-body system through entanglement.
The many-body system then evolves unitarily, and we observe how the locally encoded information is scrambled over the entire many-body system.
We note that temporal TMI has been investigated by using the channel-state duality in Ref.~\cite{Hosur2016},
while we here calculate instantaneous TMI, 
with which we can study the role of initial states. 

By studying quantum spin chains such as the XXX model and the transverse-field Ising (TFI) model with and without integrability breaking terms, we find that scrambling occurs (i.e., TMI becomes negative) for both the integrable and non-integrable systems for a majority of initial states. On the other hand, for a few initial states, scrambling does not occur (i.e., TMI becomes positive) for both the integrable and non-integrable cases of the XXX model. These results clarify that scrambling is an independent property of integrability of Hamiltonians. Therefore, scrambling does not straightforwardly correspond to conventional quantum chaos, making a sharp contrast to the level statistics and ETH. We remark that the relationship between integrability and ballistic entanglement spreading has been studied~\cite{Calabrese2005,Lauchli2008,Kim2013}, while delocalization and entanglement spreading capture different aspects of information dynamics~\cite{Bohrdt2016}, as will be discussed later in detail.

We also consider the SYK model with four-body interaction of complex fermions,
and find that disorder does not lead to slow dynamics but instead makes
scrambling smoother than a clean case.
This is contrastive to the slow scrambling in many-body localized (MBL) phase of a spin chain~\cite{Pal2010,Bardarson2012,Serbyn2013,Luitz2015,Huang2016,YChen2016,Fan2016,Swingle2016}.

\textit{Setup.}
We consider either a spin-$1/2$ or a fermionic system on a lattice, which consists of small system A on a single site and a many-body system on $L$ sites (Fig.~1).
The many-body system is divided into three subsystems B, C, D, whose sizes (the numbers of the lattice sites) are respectively given by $1$, $l$, and $L-l-1$.
The lattice structure BCD is supposed to be one-dimensional for spin chains or all-connected for the SYK model.
For a single site of a spin (fermion) system, we write $\ket{0}$ as the spin-up (particle-occupied) state, and $\ket{1}$ as the spin-down (particle-empty) state.
In any case, a single qubit is on a single site.

We first prepare a product state
\begin{align}
\frac{1}{\sqrt{2}}
(\ket{0}_\rA+\ket{1}_\rA)\otimes \ket{\Xi}_{\rB\rC\rD},
\end{align}
where
$\ket{\Xi}_{\rB\rC\rD}$
is a product state with the state of each qubit being $\ket{0}$ or $\ket{1}$ (e.g., the N$\acute{\mathrm{e}}$el state $\ket{0}\ket{1}\ket{0}\cdots\ket{0}\ket{1}$ or 
the all-up state $\ket{0}\ket{0}\cdots\ket{0}$, etc).
We then apply the CNOT gate on the state (1),
where the control qubit is A and the target qubit is B.
By this CNOT gate,
information about A is locally encoded in B through entanglement.
Then, only BCD 
obeys a unitary time evolution with a Hamiltonian.
We calculate the time dependence of TMI between A, B, C, which characterizes scrambling of the information about A that was initially encoded in B.
We note that the foregoing setup is associated with a thought experiment that one of qubits of an EPR pair is thrown into a black hole and then scrambled~\cite{Hayden2007}.

We next consider quantum-information contents.
Let X, Y, Z be subregions of the lattice (i.e., subsets of the lattice sites).
The bipartite mutual information (BMI) is defined as 
$I_2(\rX:\rY):=S_\rX+S_\rY-S_{\rX\rY}$,
where $S_\rX:=\tr_\rX[-\hat{\rho}_\rX\ln\hat{\rho}_\rX]$ is the von Neumann entropy of a reduced density operator $\hat{\rho}_\rX:=\tr_{\rX^\mathrm{c}}[\hat{\rho}]$ with $\rX^{\mathrm{c}}$ being the complemental set of X. 
Then, TMI is defined by~\cite{Hosur2016,Cerf1998} 
\begin{align}
I_3(\rX:\rY:\rZ)
&:=
I_2(\rX:\rY)+I_2(\rX:\rZ)-I_2(\rX:\rY\rZ).
\end{align}
Here, TMI is negative when
$I_2(\rX:\rY)+I_2(\rX:\rZ) < I_2(\rX:\rY\rZ)$, which implies that
information about X stored in composite YZ
is larger than 
the sum of the amounts of information that Y and Z have individually;
information about X is delocalized to Y and Z in such a case.

To illustrate the meaning of TMI,
let us consider three classical bits $x,y,z$ and the following situations:
(i)
$I_3(x:y:z)=-\ln 2$, if $x=y\oplus z$ and $y,z$ are independent and random, where $\oplus$ describes the binary sum.
In this case, neither $y$ nor $z$ is individually correlated with $x$,
but composite $yz$ is maximally correlated with $x$.
(ii)
$I_3(x:y:z)=0$, if $x,y$, and $z$ are all independent and random.
In this case, there is not any correlation between  $x,y$, and $z$.
(iii)
$I_3(x:y:z)=\ln 2$, if $x=y=z$ and $x$ is random.
In this case, the three bits form the maximum three-body correlation.

While the above examples are classical,
a similar argument applies to quantum situations.
In fact, TMI can be utilized to characterize nonlocal and long-ranged entanglement in topological orders~\cite{Kitaev2006}.

\textit{Scrambling and integrability.} 
We now discuss scrambling in the XXX model in one dimension with and without an integrability breaking term.
The Hamiltonian is given by
\begin{align}
\label{eq:Hamiltonian}
\hat{H}_{\mathrm{spin}}
:=&
\sum_{\langle i,j\rangle} J 
\mbox{\boldmath $\sigma$}_i \cdot \mbox{\boldmath $\sigma$}_j
+
\sum_{\langle\langle i,j\rangle\rangle} J^\prime 
\mbox{\boldmath $\sigma$}_i \cdot \mbox{\boldmath $\sigma$}_j,
\end{align}
where $i$ and $j$ are indices of sites,
and 
$\langle i,j\rangle$ and $\langle\langle i,j\rangle\rangle$
mean that $i$ and $j$ run within nearest neighbor (n.n.) and next nearest neighbor (n.n.n.), respectively. 
The Pauli matrices for a spin are written as $\sigma_i^\alpha$ ($\alpha=x,y,z$), and we define
$\mbox{\boldmath $\sigma$}_i:=(\sigma_i^x,\sigma_i^y,\sigma_i^z)$.
Let $J>0$. This model is integrable if $J^\prime=0$, while is non-integrable if $J^\prime >0$.

We first consider a non-integrable case with $J^\prime=0.8 J$ in Eq.~(3),
where the parameters are taken so that the level statistics is the Wigner-Dyson distribution~\cite{Santos2010}, implying that the system is fully chaotic in the sense of conventional quantum chaos.
Figure 2 shows the time dependence of TMI with the initial state being (a) N$\acute{\mathrm{e}}$el or (b) all-up, along with BMI $I_2(\rA:\rB\rC)$ (inset).
At initial time $t=0$,
BMI is given by $2\ln 2$ because of the entanglement between A and $\rB$.
As time increases, BMI decays for all the cases.
In (a), the decay is much smoother, where BMI saturates at zero for $l=1$ and at $\ln 2$ for $l=L/2-1$.
These are consistent with the behaviors of TMI as discussed below.

Figure 2 (a) shows that TMI becomes negative for the initial N$\acute{\mathrm{e}}$el state, implying scrambling.
For $l=1$,
TMI decreases from zero, goes through a minima, and gradually returns to zero.
This means that information is scrambled inside ABC in a short time regime,
and then completely disappears from BC in a longer time regime.
For $l=L/2-1$, TMI monotonically decreases and saturates at a negative value.
This means that information is scrambled but is not totally lost from BC even in a long time regime. 
These results are consistent with the behaviors of BMI.

On the other hand,
as shown in Fig.~2 (b),
TMI is positive when the initial state is all-up.
In this case, information is not scrambled, but a three-body correlation forms among A, $\rB$, and C.
On the other hand, as shown in Supplemental Material, entanglement spreads ballistically even in this case. This clarifies that what TMI characterizes is delocalization of quantum information, rather than ballistic spreading of entanglement.

We next discuss the integrable case with $J^\prime=0$ in Eq.~(3).
Figure 3 shows the time dependence of TMI for the initial state being (a) N$\acute{\mathrm{e}}$el or (b) all-up.
The qualitative behavior of TMI is similar to the non-integrable case;
scrambling occurs in (a) but does not in (b).
We do not observe recurrence induced by integrability, because our system size is sufficiently large.

We therefore conclude that scrambling occurs independently of integrability. We note that the time range of our numerical simulation is sufficiently long to see the role of non-integrability. In fact, the level spacing at the peak of the Wigner-Dyson distribution corresponds to $Jt\simeq 10^3$ in our non-integrable model~\cite{MehtaTextBook}. We also note that our numerical simulation is not restricted to the low-energy states which can be effectively described by the integrable field theory~\cite{SamajTextBook}.

To study the initial-state dependence of scrambling more systematically,
we calculated the XXX model with all possible product states $\ket{\Xi}_{\rB\rC\rD}$.
We label $2^L$ product states by bit sequences from $\ket{000\cdots0}$ to $\ket{111\cdots1}$. 
Figure 4 shows the initial-state dependence of the maximum and minimum values of TMI in $0\leq Jt<10^5$, written as $I_3^{\mathrm{max}}$ and $I_3^{\mathrm{min}}$ respectively,  for (a) non-integrable and (b) integrable cases.
The horizontal axis shows the labels of $\ket{\Xi}_{\rB\rC\rD}$ in decimal.
We see that scrambling occurs ($I_3^{\mathrm{min}}<0$) for most of the initial states.

On the other hand,
there are only four initial states with which scrambling does not occur ($I_3^{\mathrm{min}}=0$) for both of (a) and (b).
These four states are 
$\ket{0}\ket{0}\ket{0}\cdots\ket{0}$,
$\ket{1}\ket{0}\ket{0}\cdots\ket{0}$,
$\ket{0}\ket{1}\ket{1}\cdots\ket{1}$, and
$\ket{1}\ket{1}\ket{1}\cdots\ket{1}$.
The reason why these four states are exceptional is that the Hamiltonian (\ref{eq:Hamiltonian}) conserves the total magnetization in the $z$ direction. This confines the dynamics into a much smaller subspace of the Hilbert space, which leads to the absence of scrambling.

We have also calculated TMI for the TFI model with and without an integrability breaking term. The numerical results are shown in Supplemental Material, where scrambling occurs for both the integrable and non-integrable cases and for all of the initial states. The reason why scrambling occurs for the initial all-up state is that the total magnetization in the $z$ direction is no longer conserved in the TFI model, and therefore quantum information is mixed up in a huge subspace.

\textit{Sachdev-Ye-Kitaev model.}
We next consider the SYK model~\cite{Sachdev1993,Kitaev2015,Sachdev2015,Michel2016,Fu2016,Maldacena2016SYK,Danshita2016,Cotler2016,Jian2017} with complex fermions:
\begin{align}
\hat{H}_{\mathrm{SYK}}
:=
\frac{1}{(2L)^{3/2}}
\sum_{i,j,k,l}
J_{ij;kl}
c^\dag_i c^\dag_j c_k c_l,
\end{align}
where $c^\dag_i(c_i)$ is the creation (annihilation) operator of a fermion at site $i$.
The coupling in the SYK model is all-to-all and four-body (4-local),
and random:
$J_{ij;kl}$ is sampled from the complex Gaussian distribution
with variance $J^2$, satisfying $J_{ij;kl}=-J_{ji;kl}=-J_{ij;lk}=J^*_{lk;ji}$.
We also consider a clean SYK model without disorder (i.e., $J_{ij;kl}\equiv J$ for $i>j$, $k>l$) in order to clarify the role of disorder.

Figure 5 (a) shows the time dependence of TMI for the SYK model with the random coupling, where the initial state is N$\acute{\mathrm{e}}$el (i.e., fermions are half-filled), the ensemble average is taken over $16$ samples, and the error bars represent the standard deviations over the samples.
In particular, TMI for $l=L/2-1$ monotonically decreases to a negative steady value.
This smooth decrease is contrastive to the case of the clean SYK model shown in Fig.~5 (b), where scrambling occurs but TMI exhibits large temporal fluctuations.
We note that, for typical disordered cases that are also shown in Fig.~5 (b),
scrambling is smoother even without taking the ensemble average.
Therefore, disorder enhances scrambling in the case of the fermionic SYK model,
as opposed to the case of MBL of spin chains (see Supplemental Material for our numerical results on TMI in the MBL phase).

We also calculated the case of the initial all-up state (i.e., $|0\rangle \cdots|0\rangle$), where TMI is positive and scrambling does not occur for both the disordered and clean cases (see Supplemental Material). This is a consequence of the conservation of the fermion number, as is the case for the XXX model with the conservation of the initial magnetization.

\begin{table}[t]
\begin{center}
\scalebox{0.8}{
  \begin{tabular}{|c|c|c|} \hline
  	& Scrambled ($I_3<0$) & Not scrambled ($I_3>0$) \\ \hline
	Non-integrable & XXX+$J^\prime$ (N$\acute{\mathrm{e}}$el) & XXX+$J^\prime$ (all-up)  \\
	 & TFI+$h_z$ (N$\acute{\mathrm{e}}$el, all-up) &  \\ 	 \hline
	Integrable & XXX (N$\acute{\mathrm{e}}$el) & XXX (all-up) \\
	 & TFI (N$\acute{\mathrm{e}}$el, all-up) &  \\ 
	 & Clean SYK (N$\acute{\mathrm{e}}$el) & Clean SYK (all-up)  \\ \hline
	Disordered & MBL (N$\acute{\mathrm{e}}$el) & MBL (all-up)  \\
	 & Disordered SYK (N$\acute{\mathrm{e}}$el) & Disordered SYK (all-up)  \\ \hline
  \end{tabular}
  }
  \end{center}
  \caption{Summary of our numerical results.}
\end{table}

\textit{Concluding remarks.}
We have systematically investigated scrambling dynamics of quantum information in isolated quantum many-body systems, where we have adopted TMI as a measure of scrambling. We summarize the foregoing numerical results in Table I. We have observed that scrambling occurs independently of integrability, where an overwhelming majority of initial states exhibit scrambling for both the integrable and non-integrable cases. 
Although the connection between TMI and scrambling has already been established in previous works~\cite{Hosur2016,Cerf1998},
our work has newly revealed that scrambling is a separate concept from conventional quantum chaos.

We have also investigated the SYK model. We have found that disorder makes scrambling smoother in the SYK model, which is contrastive to the case of the MBL spin chain. We postpone more detailed analysis of the origin of this feature of the SYK model~\cite{IyodaKatsuraSagawaPrep}. Here we only note that the clean SYK model is integrable as shown in Table I~\cite{IyodaKatsuraSagawaPrep}.

We remark that experimental realizations of the SYK model have been theoretically proposed with ultracold atoms~\cite{Danshita2016} and a solid state device~\cite{Pikulin2017}. Furthermore, OTOC has experimentally been measured with trapped ions~\cite{Garttner2017}. By using such state-of-the-art quantum technologies, scrambling dynamics of quantum many-body systems can be investigated, and our results can be experimentally tested, which is a future issue.

T. S. is grateful to M. Rigol for a valuable discussion.
E.I. and T.S. are supported by JSPS KAKENHI Grant Number JP16H02211.
E.I. is also supported by JSPS KAKENHI Grant Number 15K20944.
T.S. is also supported by JSPS KAKENHI Grant Number JP25103003.


\begin{figure}[t]
\begin{center}
\includegraphics[width=0.6\linewidth]{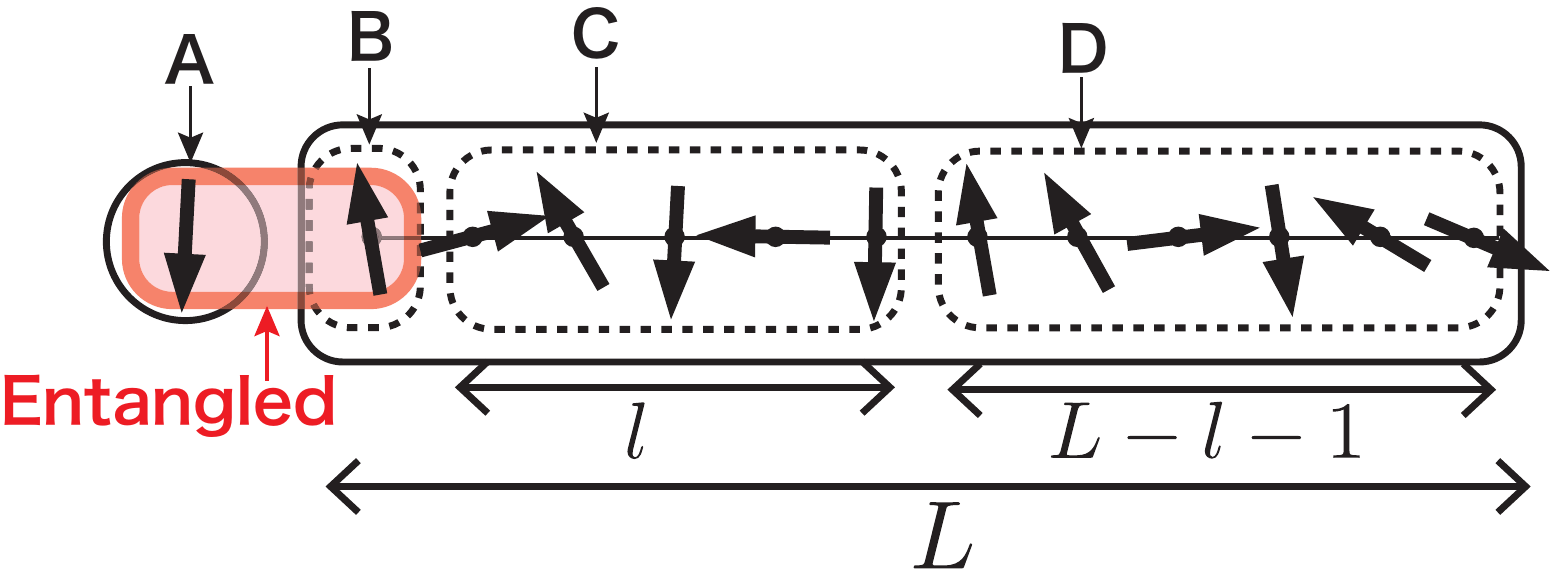}
\end{center}
\label{Main_fig1}
\caption{
(color online).
Schematics of our setup.
Initially, qubit A is maximally entangled with qubit B,
while C and D are not correlated with A, B.
Then BCD evolves unitarily with a Hamiltonian that is either integrable or non-integrable,
and either clean or disordered.
We calculate real-time dynamics of TMI between A, B, C.
}
\end{figure}

\begin{figure}[t]
\begin{center}
\includegraphics[width=0.6\linewidth]{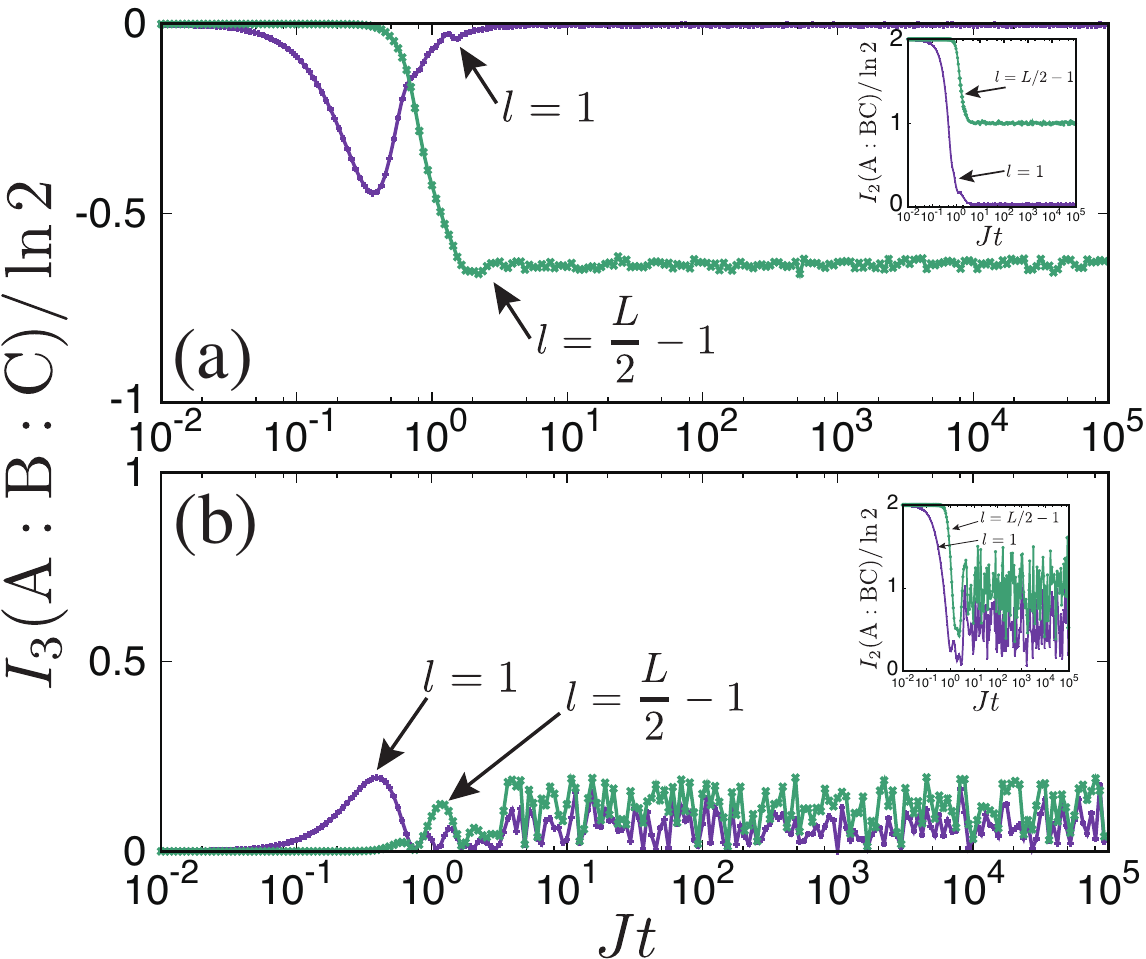}
\end{center}
\label{Main_fig2}
\caption{
(color online).
Time dependence of TMI for the non-integrable XXX model with parameters $L=14, J^\prime=0.8J$, and $l=1$ or $L/2-1$.
The initial state is (a) N$\acute{\mathrm{e}}$el or (b) all-up.
(Inset) Time dependence of BMI with the same parameters.
}
\end{figure}
\begin{figure}[t]
\begin{center}
\includegraphics[width=0.6\linewidth]{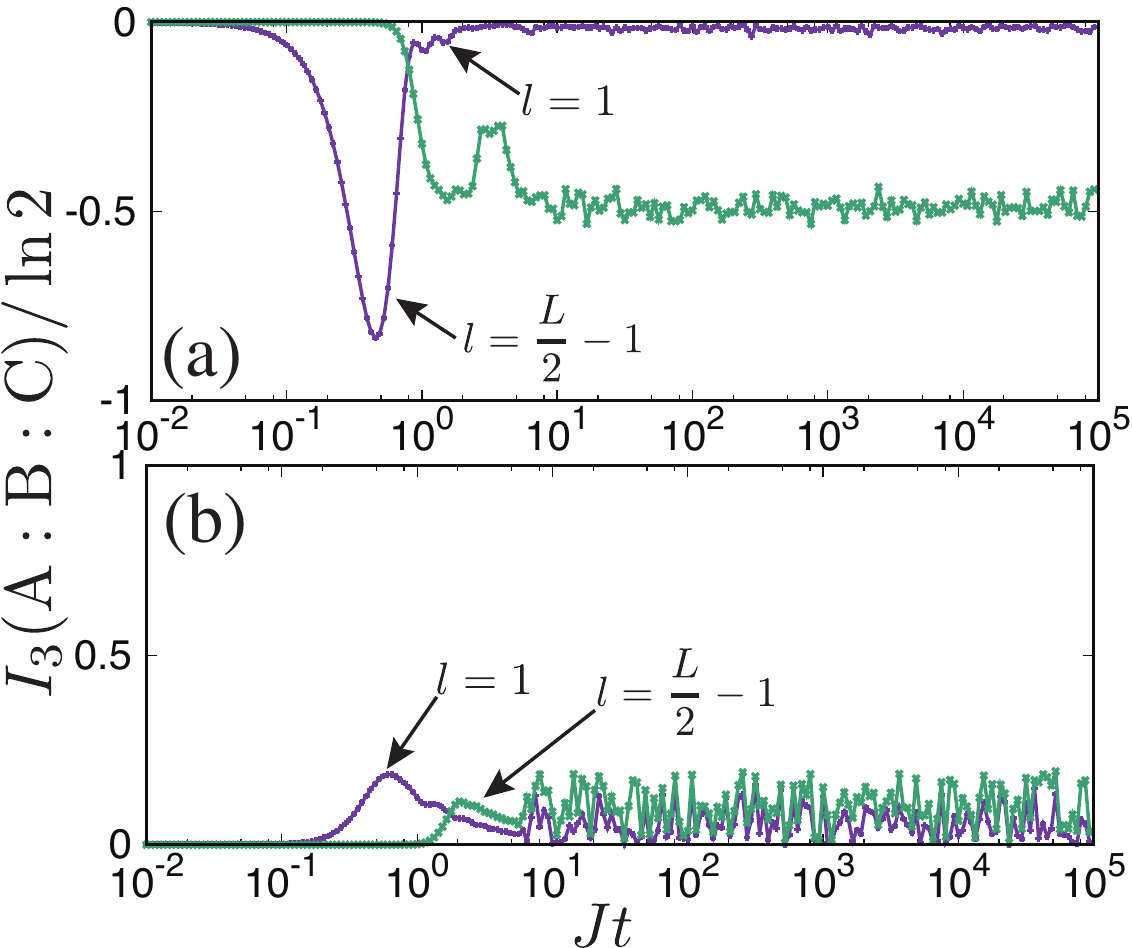}
\end{center}
\label{Main_fig3}
\caption{
(color online).
Time dependence of TMI for the integrable XXX model with parameters $L=14, J^\prime=0$, and $l=1$ or $L/2-1$.
The initial state is (a) N$\acute{\mathrm{e}}$el or (b) all-up.
}
\end{figure}

\begin{figure}[t]
\begin{center}
\includegraphics[width=0.6\linewidth]{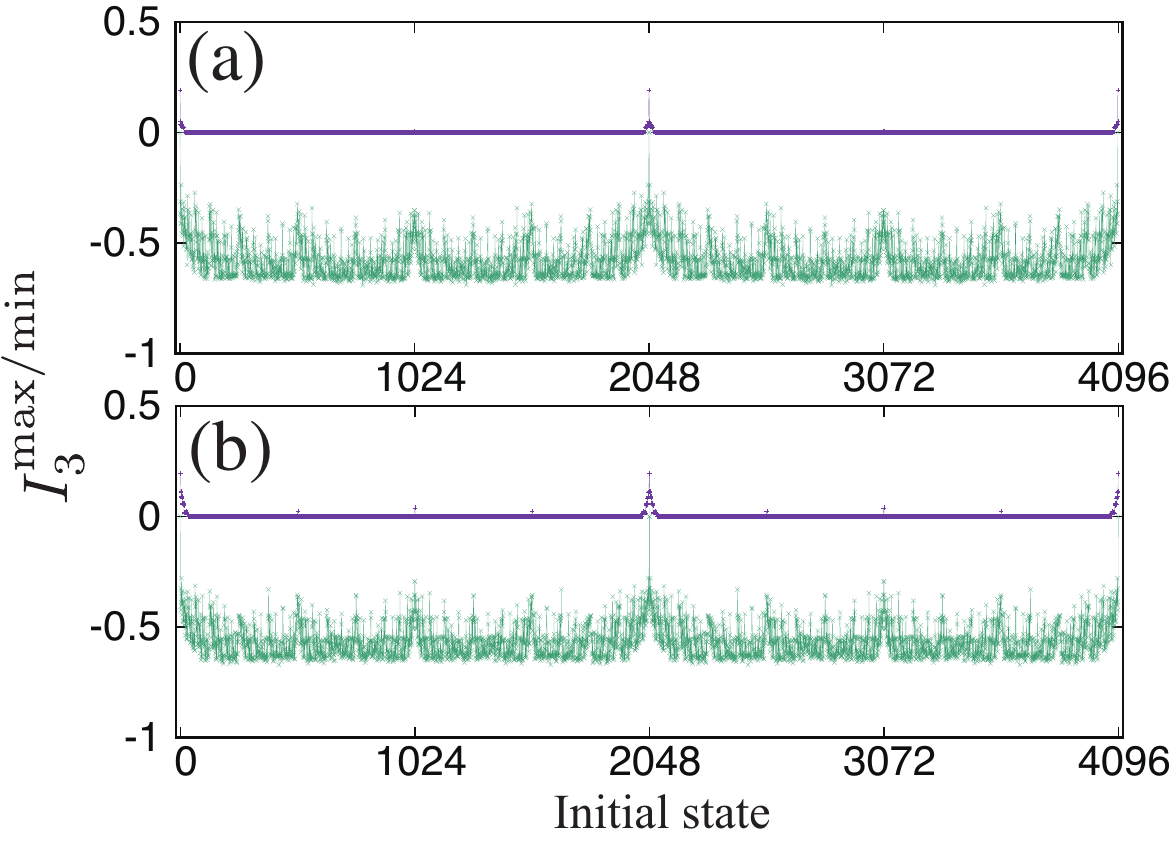}
\end{center}
\label{Main_fig4}
\caption{
(color online).
Initial-state dependence of the maximum (purple) and the minimum (green) values of TMI for the XXX model with parameters $L=12$, $l=L/2-1$.
(a) Non-integrable case ($J^\prime=0.8J$), (b) integrable case ($J^\prime=0$).
}
\end{figure}

\begin{figure}[t]
\begin{center}
\includegraphics[width=0.6\linewidth]{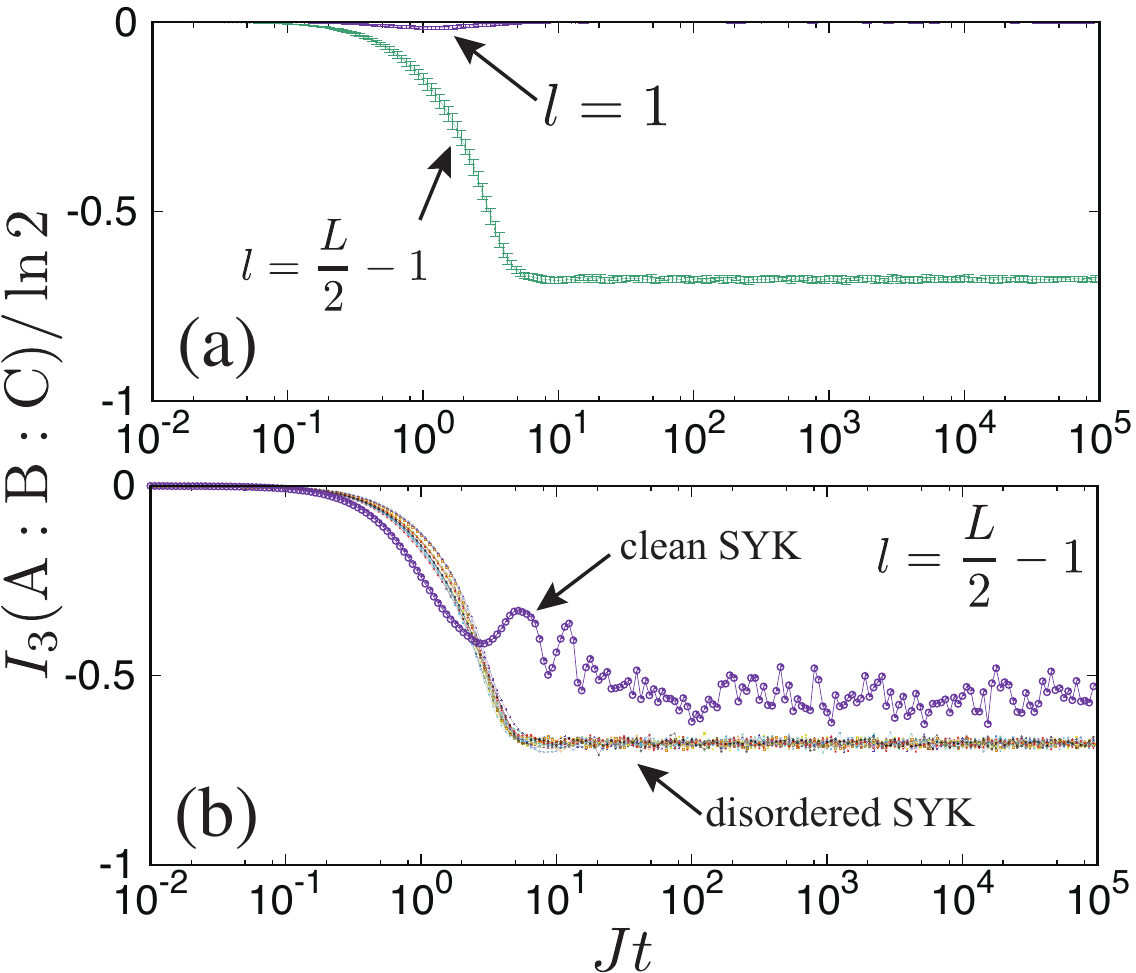}
\end{center}
\label{Main_fig6}
\caption{
(color online).
Time dependence of TMI for the SYK model with parameters $L=14$, and $l=1$ or $L/2-1$.
The initial state is the N$\acute{\mathrm{e}}$el state.
(a) Disordered SYK model. 
(b) Clean SYK model and typical samples of the disordered SYK model.
}
\end{figure}

\clearpage

{\Large Supplemental Material: Scrambling of Quantum Information in Quantum Many-Body Systems}

\

{\large Eiki Iyoda$^{1}$, Takahiro Sagawa$^{1}$}

\

[1] Department of Applied Physics, The University of Tokyo, 7-3-1 Hongo, Bunkyo-ku, Tokyo 113-8656, Japan


\vspace{10mm}

\def\theequation{S\arabic{equation}}
\def\thefigure{S\arabic{figure}}
\setcounter{equation}{0}
\setcounter{page}{1}
\setcounter{figure}{0}

\maketitle

In this Supplemental Material, we show supplemental numerical results on the bipartite mutual information (BMI) and the tripartite mutual information (TMI).
Some of the following results are mentioned in the main text.

\section{BMI for XXX model}

\begin{figure}[h]
\begin{center}
\includegraphics[width=0.9\linewidth]{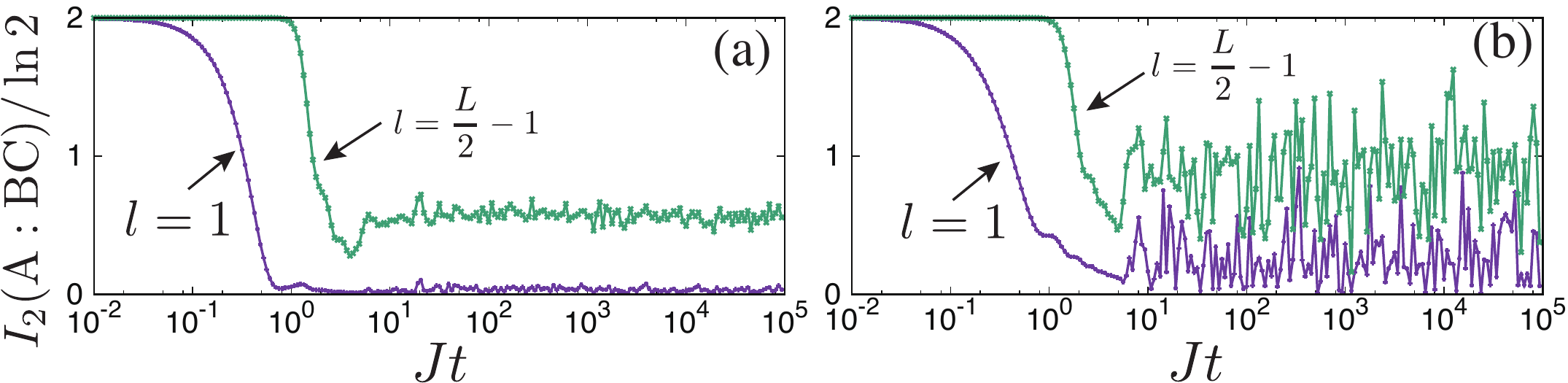}
\end{center}
\caption{
Time dependence of BMI for the integrable XXX model with parameters $L=14$ and $J^\prime=0$. 
The initial state is (a) N$\acute{\mathrm{e}}$el or (b) all-up.
}
\label{Suppl_fig1}
\end{figure}

Figure \ref{Suppl_fig1} shows the time dependence of BMI for the integrable XXX model.

\begin{figure}[h]
\begin{center}
\includegraphics[width=0.9\linewidth]{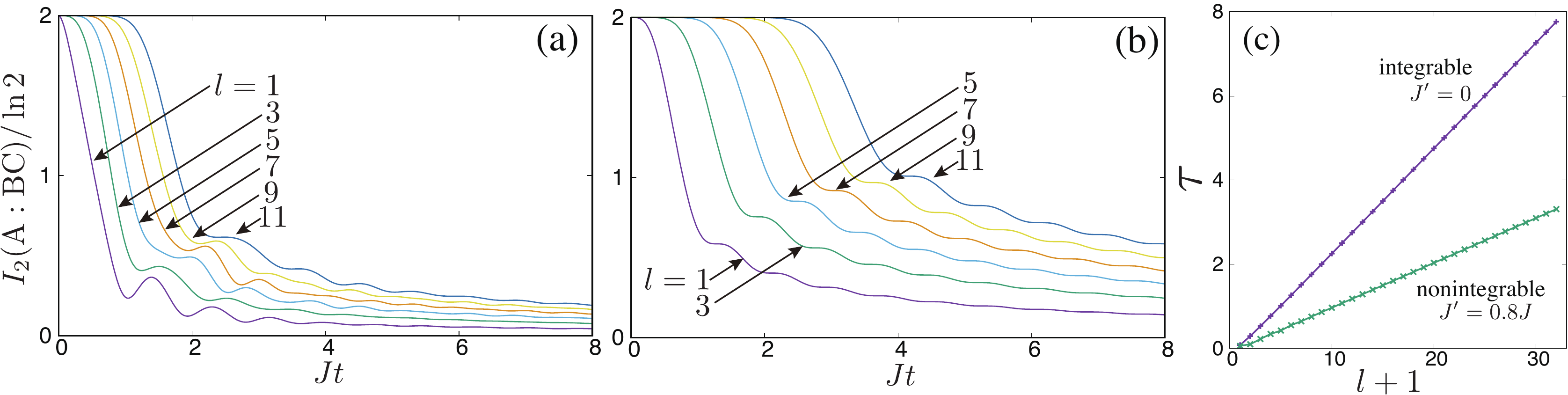}
\end{center}
\caption{
Time dependence of BMI for (a) non-integrable and (b) integrable XXX models.
The system size is $L=128$, and the initial state is all-up.
(c) Decay time $\tau$ versus subsystem size $l$.
}
\label{Suppl_figA}
\end{figure}

\begin{figure}[h]
\begin{center}
\includegraphics[width=0.7\linewidth]{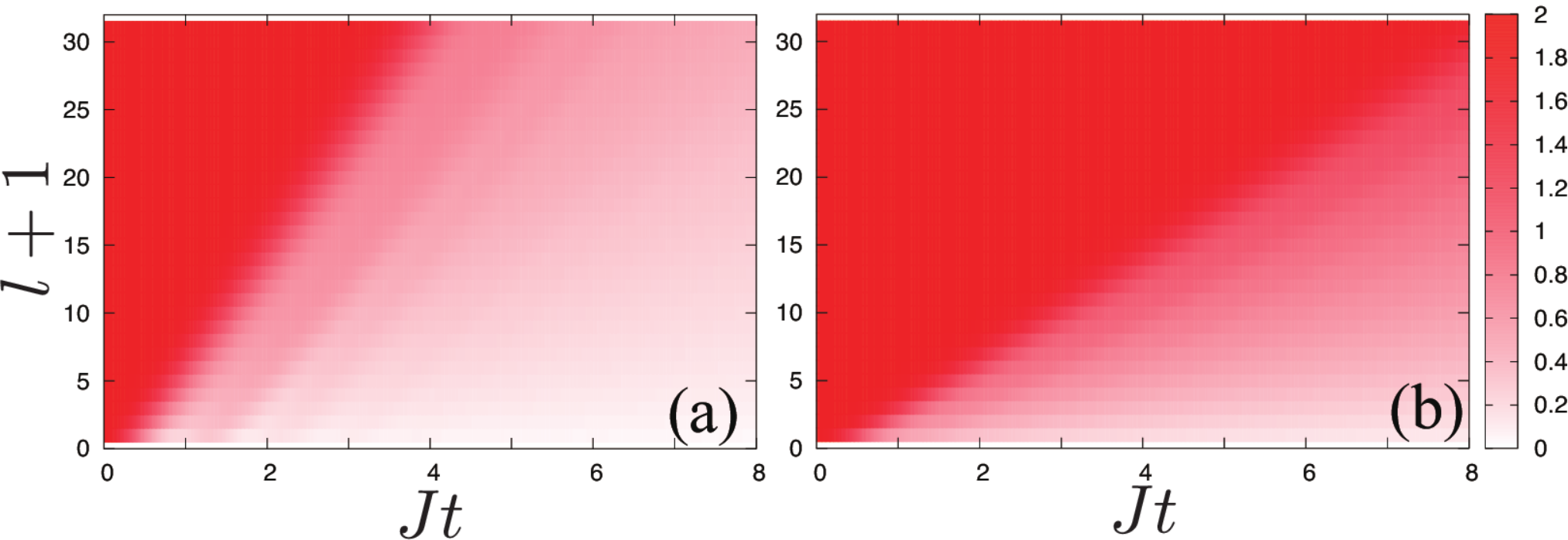}
\end{center}
\caption{
BMI versus time $t$ and subsystem size $l$ for (a) non-integrable and (b) integrable XXX models. The system size is $L=128$ and the initial state is all-up.
}
\label{Suppl_figB}
\end{figure}

In addition, we show numerical results on ballistic entanglement spreading for the integrable and non-integrable XXX models. In particular, we consider the initial all-up state, where scrambling does not occur. In this case, we can numerically access a much larger size ($L=128$), because the total magnetization is conserved in the XXX model.
Figures~\ref{Suppl_figA} (a) and (b) show the time dependence of BMI for several $l$.
Figure~\ref{Suppl_figA} (c) shows the decay time $\tau$,
at which $I_2(\rA:\rB\rC)$ becomes $a\ln 2$ ($0<a<2$) for the first time.
The constant $a$ can be arbitrarily chosen and we here set $a=1.9$.
Figure~\ref{Suppl_figA} (c) clearly shows the linear dependence of $\tau$ on $l$,
which implies that entanglement spreads ballistically. 
Figure \ref{Suppl_figB} shows BMI versus time and $l$, from which we again see the ballistic entanglement spreading.

\section{Transverse field Ising model}

The Hamiltonian of the transverse field Ising (TFI) model is given by
\begin{align}
\hat{H}_{\mathrm{TFI}}
:=
\sum_{\langle i,j\rangle}
J \sigma_i^z \sigma_j^z
+
\sum_i h_x \sigma_i^x
+
\sum_i h_z \sigma_i^z.
\end{align}
The transverse and longitudinal magnetic fields are $h_x=J$ and $h_z=0$ for an integrable case,
and are $h_x=2.1J$ and $h_z=1.1J$ for a non-integrable case.

\begin{figure}[h]
\begin{center}
\includegraphics[width=0.9\linewidth]{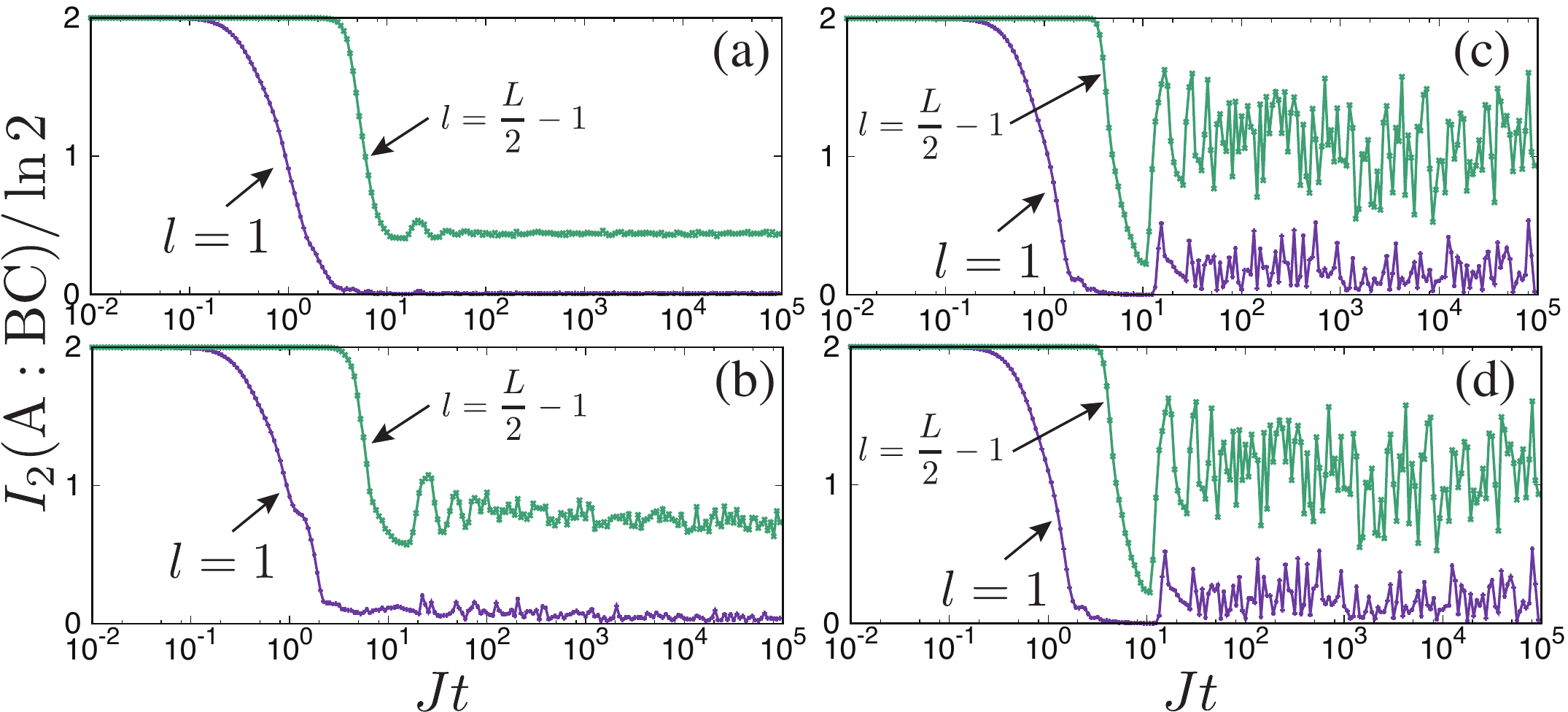}
\end{center}
\caption{
Time dependence of BMI for the TFI model with $L=14$.
The integrability and the initial state are
(a) non-integrable/N$\acute{\mathrm{e}}$el,
(b) non-integrable/all-up,
(c) integrable/N$\acute{\mathrm{e}}$el, and
(d) integrable/all-up.
}
\label{Suppl_fig2}
\end{figure}

\begin{figure}[h]
\begin{center}
\includegraphics[width=0.9\linewidth]{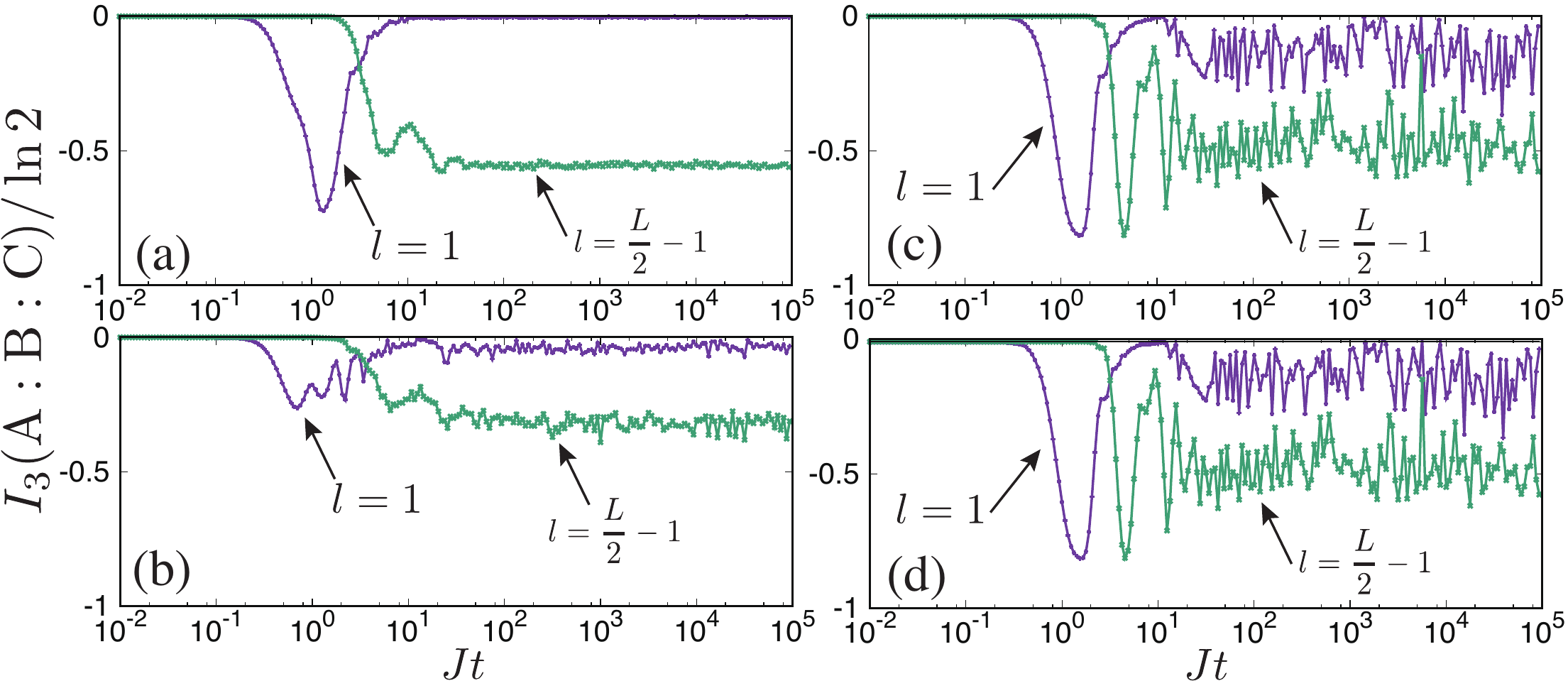}
\end{center}
\caption{
Time dependence of TMI for TFI with $L=14$.
The integrability and the initial state are
(a) non-integrable/N$\acute{\mathrm{e}}$el,
(b) non-integrable/all-up,
(c) integrable/N$\acute{\mathrm{e}}$el, and
(d) integrable/all-up.
}
\label{Suppl_fig3}
\end{figure}

Figures~\ref{Suppl_fig2} and \ref{Suppl_fig3} show the time dependence of BMI and TMI for the TFI model, respectively.
Figure \ref{Suppl_fig3} shows that scrambling occurs independently of integrability or the initial state.

\begin{figure}[h]
\begin{center}
\includegraphics[width=0.9\linewidth]{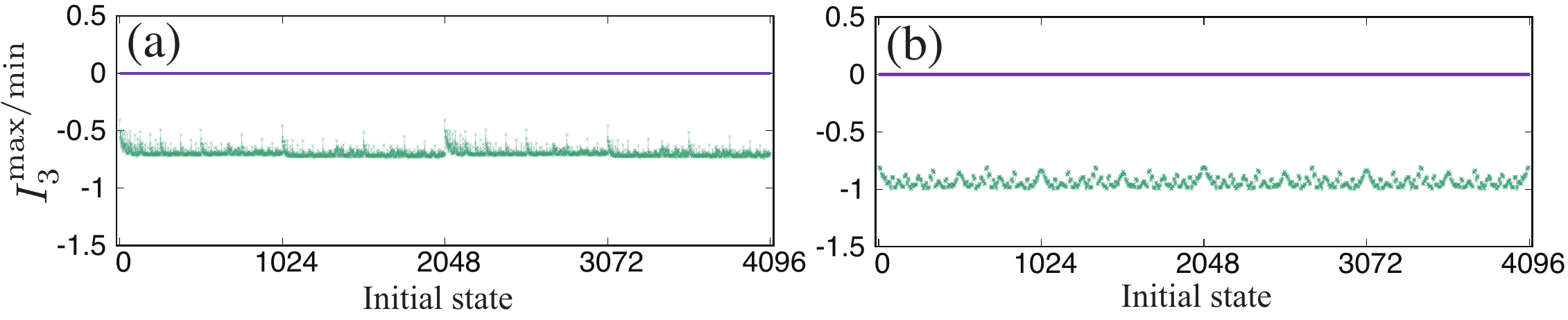}
\end{center}
\caption{
Initial-state dependence of the maximum (purple) and the minimum (green) values of TMI for the TFI model 
with parameters $L=12$ and $l=L/2-1$. (a) Non-integrable case ($h_x=2.1J$ and $h_z=1.1J$), (b) integrable case ($h_x=J$ and $h_z=0$).
}
\label{Suppl_figC}
\end{figure}

Figure \ref{Suppl_figC} shows that the initial-state dependence of $I_3^\mathrm{max}$ and $I_3^\mathrm{min}$,
where the initial states are represented by decimal from $|000\cdots 0\rangle$ to $|111\cdots 1\rangle$, as in Fig.~4 of the main text.
Scrambling occurs for both (a) non-integrable and (b) integrable cases for all of the initial states without exception.

\section{Disordered systems}
\subsection{MBL systems}

The Hamiltonian of disordered spin chains is given by
\begin{align}
\hat{H}
:=&
\sum_{\langle i,j\rangle} J 
\mbox{\boldmath $\sigma$}_i \cdot \mbox{\boldmath $\sigma$}_j
+
\sum_{i} h_i \sigma^z_i,
\end{align}
where $h_i$ is a random magnetic field and is generated uniformly from $[-h,h]$ ($h\geq 0$).

\begin{figure}[h]
\begin{center}
\includegraphics[width=0.9\linewidth]{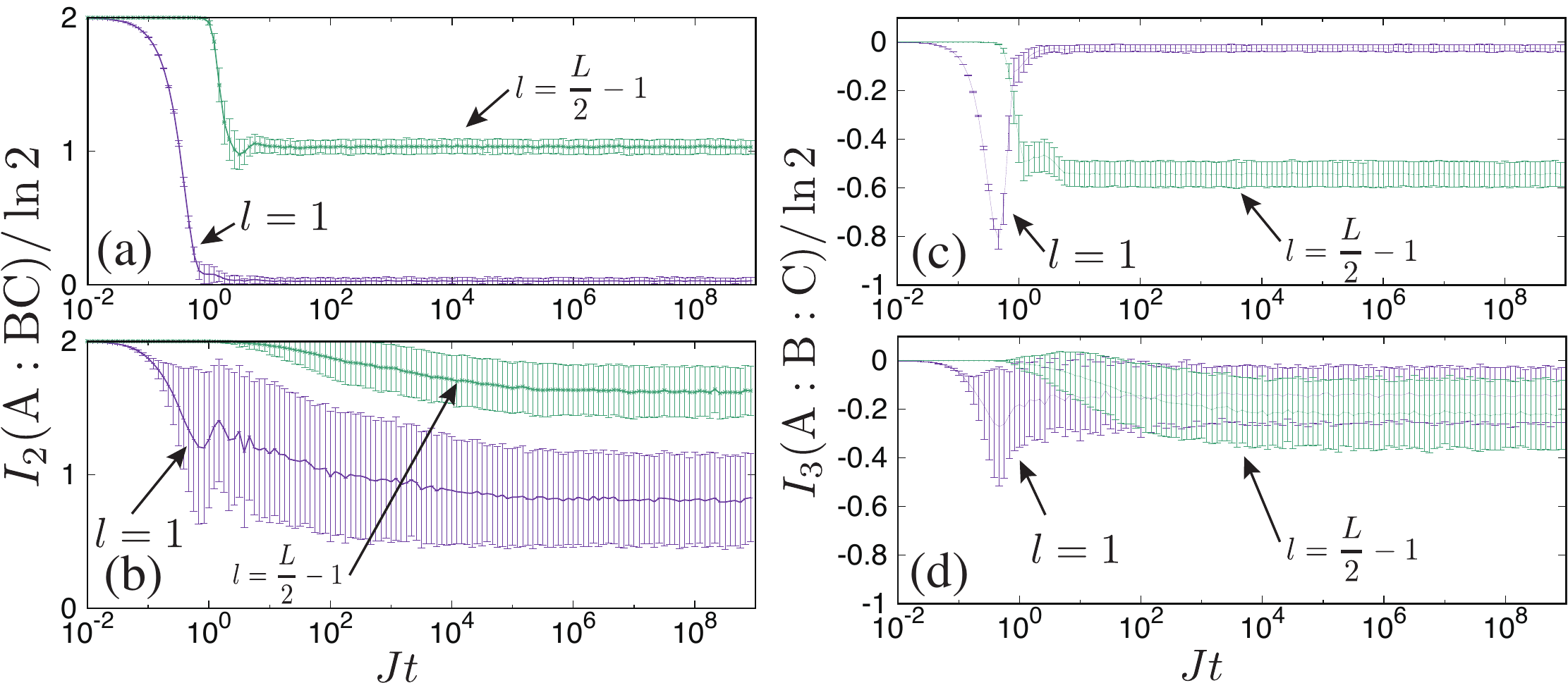}
\end{center}
\caption{
Time dependence of BMI 
for the disordered XXX model with the initial N$\acute{\mathrm{e}}$el state with $L=12$.
The ensemble average is taken over $128$ samples.
The informational content and the phase are
(a) BMI/ergodic ($h=J$),
(b) BMI/MBL ($h=10J$),
(c) TMI/ergodic ($h=J$), and
(d) TMI/MBL ($h=10J$).
}
\label{Suppl_fig4}
\end{figure}

Figure~\ref{Suppl_fig4} shows the time dependence of BMI and TMI in the disordered XXX model with the initial N$\acute{\mathrm{e}}$el state. In the MBL phase (Fig.~\ref{Suppl_fig4}(b)),  BMI decays quite slowly. This is consistent with a phenomenology of local integrals of motion of MBL~[47].

\begin{figure}[h]
\begin{center}
\includegraphics[width=0.9\linewidth]{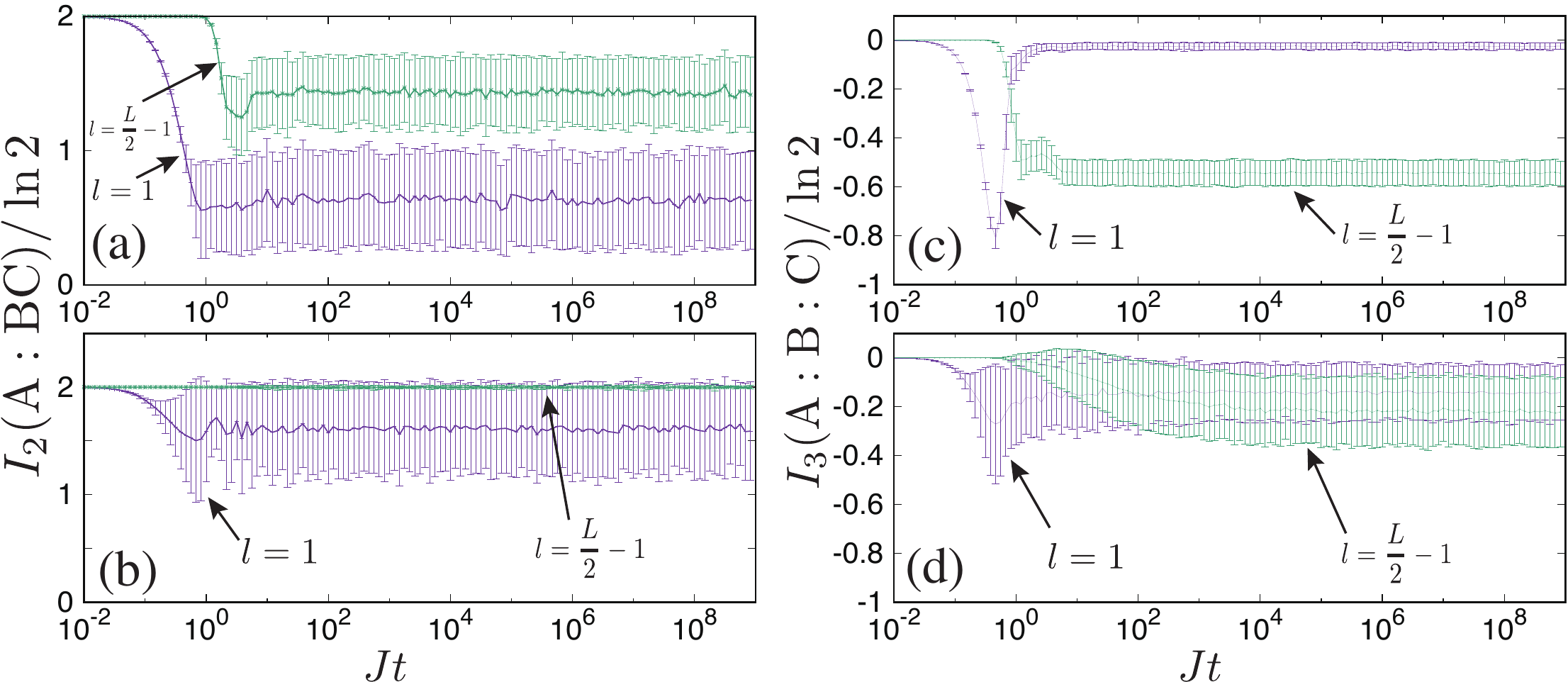}
\end{center}
\caption{
Time dependence of BMI and TMI for 
the disordered XXX model with the initial all-up state with $L=12$.
The number of samples is $128$.
The informational content and the phase are
(a) BMI/ergodic ($h=J$),
(b) BMI/MBL ($h=10J$),
(c) TMI/ergodic ($h=J$), and
(d) TMI/MBL ($h=10J$).
}
\label{Suppl_fig5}
\end{figure}

Figure~\ref{Suppl_fig5} shows the time dependence of BMI and TMI for the disordered XXX model with the initial all-up state. As shown in Figs.~\ref{Suppl_fig5} (c) and (d), scrambling does not occur both in the ergodic and MBL phases.

\subsection{SYK model}
\begin{figure}[h]
\begin{center}
\includegraphics[width=0.5\linewidth]{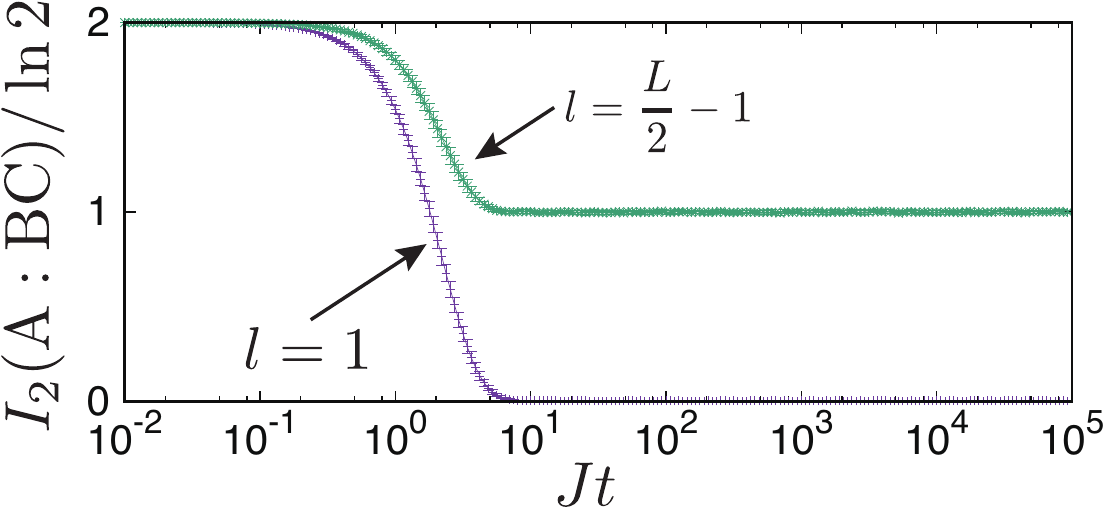}
\end{center}
\caption{
Time dependence of BMI for the disordered SYK model with $L=12$ 
and with the initial N$\acute{\mathrm{e}}$el state.
The ensemble average is taken over $16$ samples.
}
\label{Suppl_fig6}
\end{figure}

\begin{figure}[h]
\begin{center}
\includegraphics[width=0.9\linewidth]{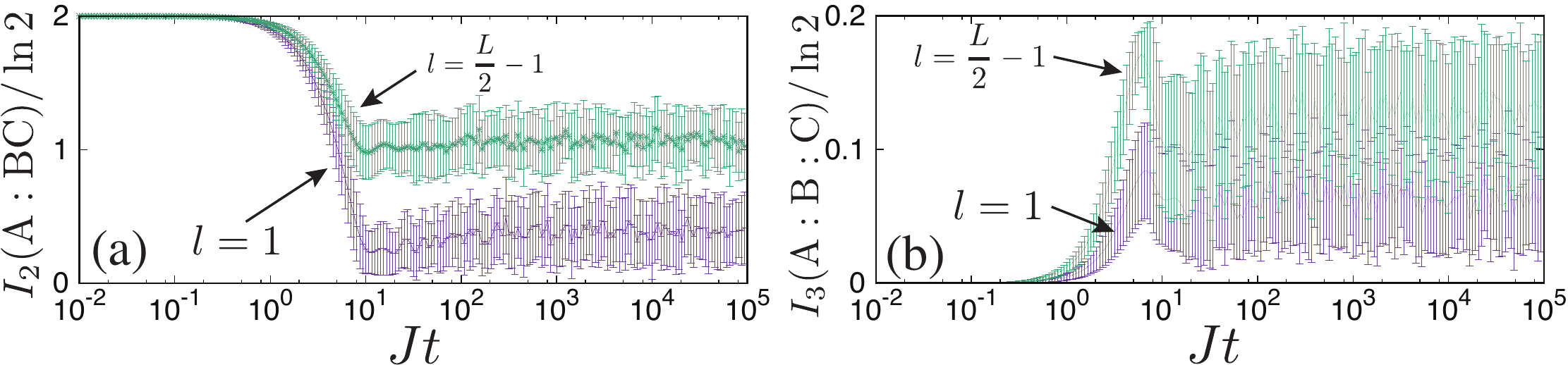}
\end{center}
\caption{
Time dependence of (a) BMI and (b) TMI for the disordered SYK model with $L=10$ and with the initial all-up state.
The ensemble average is taken over $16$ samples.
(a) BMI and (b) TMI.
}
\label{Suppl_fig7}
\end{figure}

Figure~\ref{Suppl_fig6} shows the time dependence of BMI for the disordered SYK model 
with the initial N$\acute{\mathrm{e}}$el state.
Figure~\ref{Suppl_fig7} shows the time dependence of BMI and TMI 
for the disordered SYK model with the initial all-up state.
As shown in Fig.~\ref{Suppl_fig7} (b), scrambling does not occur as TMI is positive.

\end{document}